\documentclass[sigconf]{acmart}

\AtBeginDocument{%
  \providecommand\BibTeX{{%
    \normalfont B\kern-0.5em{\scshape i\kern-0.25em b}\kern-0.8em\TeX}}}

\copyrightyear{2025}
\acmYear{2025}
\setcopyright{rightsretained}
\acmConference[IDC '25]{Interaction Design and Children}{June 23--26,
2025}{Reykjavik, Iceland}
\acmBooktitle{Interaction Design and Children (IDC '25), June 23--26, 2025,
Reykjavik, Iceland}\acmDOI{10.1145/3713043.3731498}
\acmISBN{979-8-4007-1473-3/2025/06}


\usepackage{geometry}
\usepackage{array}
\usepackage{caption}
\usepackage{booktabs}
\usepackage{graphicx}
\usepackage{subcaption}
\usepackage{tabularx}
\usepackage{colortbl}
\usepackage{xcolor}
\usepackage{float}
\definecolor{lightgray}{gray}{0.9}

\begin{document}
\title[Understanding Children's Hopes, Fears, and Evaluations of Generative AI] {``If anybody finds out you are in BIG TROUBLE'': Understanding Children's Hopes, Fears, and Evaluations of Generative AI}

\author{Aayushi Dangol}
\affiliation{%
  \institution{University of Washington}
  \city{Seattle}
  \state{WA}
  \country{USA}
}
\email{adango@uw.edu}

\author{Robert Wolfe}
\affiliation{
  \institution{University of Washington}
  \city{Seattle}
  \state{WA}
  \country{USA}
}
\email{rwolfe3@uw.edu}

\author{Daeun Yoo}
\affiliation{
  \institution{University of Washington}
  \city{Seattle}
  \state{WA}
  \country{USA}
}
\email{daeunyoo@uw.edu}

\author{Arya Thiruvillakkat}
\affiliation{
  \institution{George Mason University}
  \city{George Mason}
  \state{VA}
  \country{USA}
}
\email{athiruvillakkat@gmail.com}

\author{Ben Chickadel}
\affiliation{
  \institution{University Child Development School}
  \city{Seattle}
  \state{WA}
  \country{USA}
}
\email{benc@ucds.org}

\author{Julie A. Kientz}
\affiliation{
  \institution{University of Washington}
  \city{Seattle}
  \state{WA}
  \country{USA}
}
\email{jkientz@uw.edu}

\renewcommand{\shortauthors}{Dangol et al.}

\begin{abstract}
As generative artificial intelligence (genAI) increasingly mediates how children learn, communicate, and engage with digital content, understanding children's hopes and fears about this emerging technology is crucial. In a pilot study with 37 fifth-graders, we explored how children (ages 9–10) envision genAI and the roles they believe it should play in their daily life. Our findings reveal three key ways children envision genAI: as a companion providing guidance, a collaborator working alongside them, and a task automator that offloads responsibilities. However, alongside these hopeful views, children expressed fears about overreliance, particularly in academic settings, linking it to fears of diminished learning, disciplinary consequences, and long-term failure. This study highlights the need for child-centric AI design that balances these tensions, empowering children with the skills to critically engage with and navigate their evolving relationships with digital technologies.

\end{abstract}

\maketitle

\section{Introduction}
Generative AI (genAI) is increasingly shaping children’s digital experiences, from personalized learning tools to AI-assisted creative platforms. Yet, its role is fraught with both promise and peril. On the one hand, genAI is hailed as a revolutionary force that can democratize education, enhance creativity, and provide adaptive learning experiences \citep{han2023design, wu2023integrating, divanji2024togethertales, Dangol2025}. On the other hand, concerns arise over genAI's potential to diminish critical thinking and foster over reliance on machine-generated content \citep{alasadi2023generative, lee2025impact, dangol2025ai}. As generative AI systems become more embedded in children's lives, discussions surrounding its impact remain deeply ambivalent. However, much of the discourse about its benefits and risks is framed from an adult-centric perspective, often emphasizing technical safeguards and guidelines without incorporating children’s voices. 

While prior studies have explored children’s perceptions of AI from a hopes and fears lens, they have largely focused on embodied AI such as robots and voice assistants \citep{rubegni2022don, collyer2024s}. However, unlike embodied AI, which provides clear boundaries between human and machine interaction, genAI can be embedded into creative, educational, and social platforms in ways that blur the distinction between human and algorithmic contributions. Given children’s hopes and fears shape their interactions with technology \citep{yip2019laughing, cave2019hopes} understanding their perspectives on genAI is essential for designing child-centered AI experiences that align with their cognitive, emotional, and social needs. Thus, to understand how children conceptualize genAI's use in their everyday lives, we conducted a three-session pilot study with 37 fifth-grade students (ages 9–10) at a local elementary school in the United States. Our findings capture children's excitement about genAI’s potential to foster creativity, motivation, and play, as well as their fears on over reliance on genAI leading to diminished skills and negative academic consequences.

\section{Related Work}
While research on children's use of generative AI is emerging, early surveys suggest that children and youth are engaging with genAI at increasing rates, often surpassing adult adoption \cite{pew2025chatgpt, commonsense2025ai}. Given this trend, studies within HCI research are exploring how genAI can support children's creativity \cite{10.1145/3585088.3594495, newman2024want, resnick2024generative}, storytelling \cite{han2023design, zhang2021storydrawer, zhao2024magic,fan2024storyprompt}, and learning \cite{su2023unlocking, alasadi2023generative, qadir2023engineering, wu2023integrating}, while also examining biases within genAI models that affect children and teenagers \cite{wolfe2024representation, wolfe2024dataset, Lewis2025}. Studies of children's understanding of genAI suggest that children (ages 5–12) primarily view it as a tool for producing content \citep{kosoy2024children} and are capable of understanding and articulating the cultural implications of its use \citep{dangol2024mediating}. However, research also indicates that children tend to overestimate genAI's capabilities \citep{dietz2023theory, dangol2025mental} and without structured guidance they may struggle to discern errors, particularly when genAI presents information in a coherent yet misleading way \citep{solyst2024children, dangol2025ai}.

Thus, it is essential to examine not only how children use and understand generative AI but also what they consider important in a technological society \citep{rubegni2022don}. Our study draws on the ``hopes and fears'' model, which provides a framework for understanding user perceptions of emerging technologies within the dichotomy of optimism and apprehension \cite{cave2019hopes}. This model has been successfully applied in Child-Computer Interaction research to examine how children's hopes and fears shape their engagement with technology ~\cite{livingstone2020parenting}. For example, research by Yip et al. highlights that children often find technologies ``creepy''  when they provoke fears related to physical safety or emotional discomfort ~\cite{yip2019laughing}. Similarly, studies have explored children's hopes and fears of embodied AI such as social robots ~\cite{rubegni2022don, collyer2024s}, as well as how these perceptions influence their interactions with intelligent toys and shape parental views on such technologies ~\cite{livingstone2020parenting, mcreynolds2017toys}. Expanding this perspective beyond children and families, Smakman et al. examined how various stakeholders navigate the integration of social robots into primary education, shedding light on the broader societal concerns surrounding these technologies ~\cite{smakman2021moral}. While much of the existing discourse has focused on children's hopes and fears regarding embodied AI, prior research also suggests that the form of AI matters in shaping children's perceptions \citep{flanagan2023growing, dietz2023theory, quander2024you}. Given the non-embodied nature of generative AI, our work seeks to understand children's hopes and fears in this context. 

\section{Methods}
We conducted our pilot study in two fifth-grade classrooms at a local U.S. elementary school. A total of 37 students (ages 9–10) participated during their regular Technology class, taught by their Technology teacher. Children reported varying levels of AI use, with 76.3\% of participants having used voice assistants (\textit{e.g.}, Alexa, Siri), 68.4\% have used video game AIs, 23.7\% have engaged with chatbots, and 5.3\% having no direct AI experience. We obtained parental consent and child assent for all participants, and our university's Institutional Review Board (IRB) reviewed and approved all research related activities.

\subsection{Classroom Sessions}
\begin{figure*}
    \centering
    \includegraphics[width=0.8\linewidth]{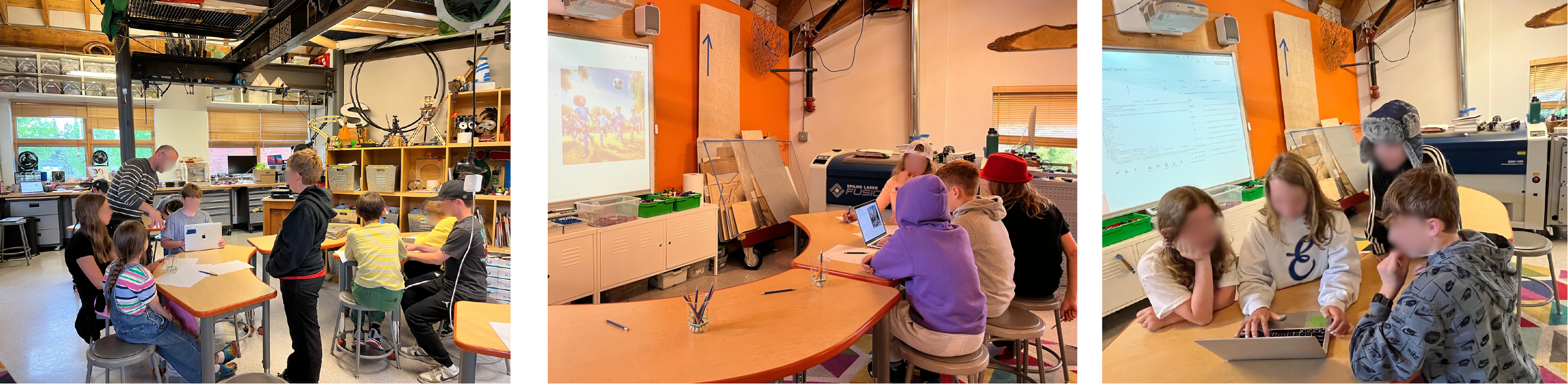}
    \caption{Fifth grade students engaging with generative AI during their Technology class.}
    \label{fig:class}
\end{figure*}

We conducted three sessions each with two fifth grade classrooms between April and May 2024 (see Figure \ref{fig:class}). Each classroom session lasted 60 minutes, consisting of Warm-Up Time (10 minutes), Activity Time (30 minutes), and Reflection Time (20 minutes). To mitigate potential power imbalances and the influence of children’s responses on each other, the teacher navigated the discussions in a way that minimized dominant behavior of any single participant and promoted collective idea generation, where contributions were built upon rather than overshadowed. 

\begin{figure*}
    \centering
    \includegraphics[width=0.8\linewidth]{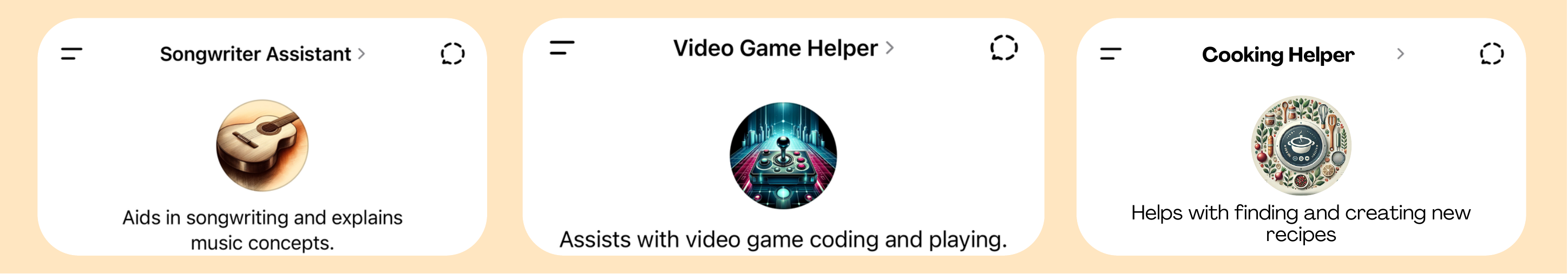}
    \caption{Examples of genAI agents used by students, assisting with songwriting, video game development, and cooking.}
    \label{fig:genAI}
\end{figure*}

In the first session, the Technology Teacher introduced children to genAI through ChatGPT-4o, demonstrating its core functionalities, including image generation, voice-based interactions, and question-answering capabilities. Afterward, students created a story and drawing, illustrating potential applications of genAI in their daily lives. Based on student narratives and drawings from Session 1, the research team developed six distinct genAI agents using the OpenAI GPT-Store\footnote{https://chatgpt.com/gpts} for students to explore in Session 2 (see Figure \ref{fig:genAI}). These agents assisted with cooking, homework, sports, games, songwriting, and theatre. Working in groups of 5–6, students took turns interacting with different genAI agents and completed a worksheet exploring situations where human judgment was preferable over AI-generated support. In Session 3, students developed their own evaluation criteria to assess the genAI agents they explored in Session 2. After establishing their evaluation framework, students revisited the six genAI agents, interacting with them again while applying their newly developed criteria. They then rated each agent using a three-tiered system: Best, Average, or Worst, based on how well genAI met their expectations in different areas. Following the evaluation process, students engaged in a group discussion to compare their ratings and reflect on the perceived benefits and drawbacks of AI assistance. 

During the classroom sessions, we collected worksheets and artifacts produced by the students, along with observational notes from the research team. We transcribed students' written responses from their worksheets and analyzed the data using a thematic analysis approach \citep{braun2022conceptual}. The authors developed inductive codes by reviewing and open-coding the data, then collaboratively discussed and refined the codes throughout the process to resolve discrepancies, reach a consensus, and organize findings into high-level themes.

\section{Findings}
\subsection{Hope for AI to Enrich Everyday Experiences} In Session 1, children expressed hope that genAI could be integrated into their daily lives, envisioning its role at varying levels of involvement: 1) as an advisor providing guidance, 2) as a collaborative assistant working alongside them, and 3) as a task automator handling responsibilities on their behalf. At the most minimal level of involvement, children framed genAI as a support system, helping them stay organized, make decisions, or sustain motivation without directly taking over tasks. For example, Mira (all names are pseudonyms) viewed genAI as a planner that could help her with time management,``\textit{I usually procrastinate, so this could help space out my work. Also, when my dad is busy, I cannot ask for help. AI can help me schedule planning to manage my time.}'' Similarly, Sophie envisioned genAI simplifying decision-making, ``\textit{Because I have a hard time cleaning my room, it would help me make decisions on what to keep and where to put things. I would follow the instructions that it gives me.}'' Children also hoped for genAI to serve as a motivator, offering encouragement as they completed their tasks. For example, Jude noted, \textit{“By picking me up when things are hard and making me laugh,} while Noah mentioned, ``\textit{every time I have to do my laundry, it helps me fold it by playing music and talks to me while I fold.}''

A step further, several children viewed genAI as a collaborator in their creative and recreational activities, working alongside them to extend their capabilities. Clara imagined genAI assisting in songwriting, ``\textit{My AI helper would make songwriting easier by suggesting relevant adjectives, lyrics, sounds, etc. It would come up with a tune to go with the lyrics whenever inspiration strikes me.}'' Similarly, Roe shared, ``\textit{When I have an audition or a play that I have to memorize lines for I have no one to practice my lines with. It would be so much easier and would be more fun when the AI can play the other character in the script. It will help me get more emotions and won't have to worry about others lines.}'' Children also saw genAI as a partner in gaming, with Owen noting, ``\textit{If I’m playing video games Zelda or Minecraft and I don’t know what to do, AI would help me identify facts or other things that appear in game that I don't know. [So] I would always know where to go and how to do it.}'' Similarly, Lena positioned AI as a playmate, stating, ``\textit{If no one is home, AI could play with or against me so I wouldn’t have to play by myself.}''

At the deepest level of involvement, some children envisioned genAI fully taking over certain responsibilities. Ava, for example, envisioned genAI completing her homework, ``\textit{I often spend a lot of my time doing my homework, but my AI helper would make this task easier by doing it for me so I can practice my hobbies and languages. I would [instead] watch a movie, videos, or make DIY projects.}'' Similarly, Ora described, \textit{``I wouldn’t have to play piano, which would free up 30 minutes of my time. I would read, paint, write, sculpt, or do homework. I would probably be less stressed.''} While earlier forms of genAI involvement were described in terms of shared participation, these examples suggest that children also see value in genAI’s ability to offload tasks that feel burdensome, allowing them to pursue activities they find more meaningful.

\subsection{Fear of Losing Learning Opportunities and Impact on Future Success} In Session 2, children shared their fears that relying on genAI for answers could undermine their ability to develop problem-solving skills, ultimately jeopardizing their learning, academic integrity, and future success. This fear was especially evident when children discussed using genAI for completing homework. As Ava stated, \textit{''Your homework won't have any wrong answers, but if the AI did your homework for you, then you won't get any practice.''}  Similarly, Max emphasized, \textit{''You would get most of your homework right but would not learn important skills and fail most tests because you would not have the AI to do it for you.''} Their fear of failing tests without genAI support highlights an implicit trust in its accuracy, despite acknowledging the drawbacks of dependence. Children also had a general belief that answer-generation without guidance from genAI could significantly hinder learning. For example, Tessa cautioned, \textit{If you don't do it yourself [and] if [AI] doesn't explain what it's doing, you don't learn, and you fall behind.''} These reflections highlight that children recognize the importance of productive struggle in learning rather than simply receiving correct answers from genAI. 

\begin{figure*} 
    \centering
    \includegraphics[width=0.8\linewidth]{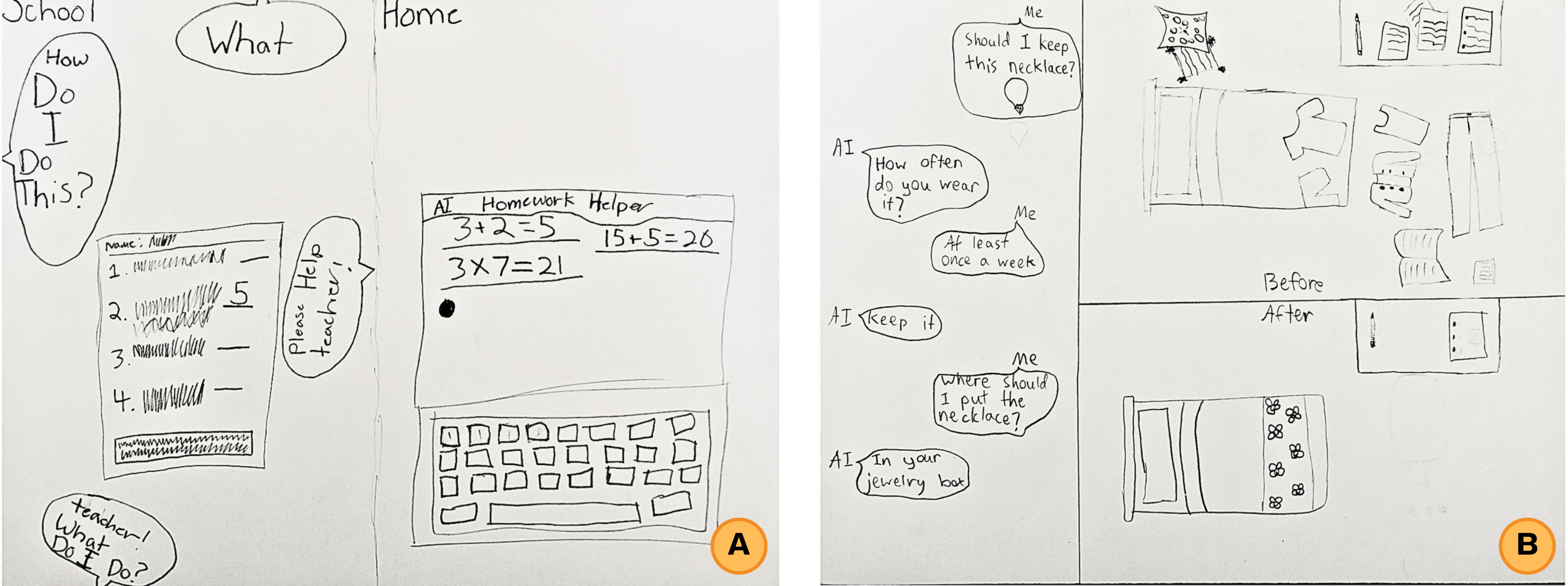}
    \caption{Children's conceptualization of generative AI as a supportive tool in their daily life. The drawings depict genAI as a homework assistant (A) and as a helper in organizing a room (B), offering suggestions on what to keep and discard.}
    \label{fig:enter-label}
\end{figure*}

Children also expressed fears about potential academic and disciplinary consequences of AI-assisted homework. For example, they noted that teachers could detect inconsistencies in students' work and that they could ``get into BIG trouble''. As Ora speculated, ``\textit{You will learn very little, and most likely will be questioned by teachers about your high homework grades but low test grades, and might have to go down a few grades--or get expelled!}'' Similarly, Harper remarked, \textit{``If you get caught, you might have to retake middle and high school.''} Children's emphasis on detection and punishment highlight fears that children view AI use in schoolwork as a high-risk behavior, underscoring the perception that genAI is more of a shortcut than a learning tool. 

Beyond immediate disciplinary concerns, children also expressed fears about the long-term consequences of genAI reliance, particularly regarding future academic and career success. Some envisioned genAI use as a slippery slope that could negatively impact their ability to succeed in college or the workplace. As Finn speculated, ``\textit{You probably won’t get into high school. Since you fail all of the tests. You won’t get a job and you will die of starvation because you did not graduate.}'' Similarly, Hazel stated, ``\textit{You will be missing out on a lot of stuff because you don’t know how to do anything, and you won’t get into some colleges.}'' Some children also framed their concerns in terms of cognitive decline, expressing fears that genAI use would make them less intelligent. For example, Sophie stated, ``\textit{You get dumb because you have no practice,}'' while Yuki reflected, ``\textit{I won’t have to do anything. I become lazy.}'' These statements highlight children's fear that dependence on genAI reduces the need for personal effort, ultimately leading to failure in life.

\subsection{Children’s Criteria for Evaluating Generative AI}
In Session 3, children developed their own evaluation criteria to assess their interactions with genAI. This process provided a tangible way for children to explore whether using genAI aligned with their hopes and fears, helping shape their understanding of genAI’s potential role in their lives. While children's individual criteria varied, common themes emerged across the group. Specifically, children consistently prioritized accuracy, speed, clarity, creativity, and social-emotional engagement as key qualities in their evaluation. However, children’s ratings for each quality varied, reflecting both enthusiasm and disappointment. For example, when evaluating genAI's creativity, Mira gave the rating ``best'' stating, ``\textit{It came up with designs I would never think off}.'' In contrast, Tessa for the same output rated it as ``worse'' noting ``\textit{Its responses were boring because it gave really standard response, did not add any cultural references, it was bland.}'' Similarly, while children expressed a strong preference for genAI to interact in a socially engaging and emotionally responsive way, they differed in their evaluations of how well it met these expectations. For example, while Ava praised genAI's friendliness, stating, ``\textit{The AI was so friendly, almost like a real person,}'' Yuki noted ``\textit{It sounded like a rad person saying stuff but could not show emotion. It was descriptive but was emotionless.}'' Children's mixed evaluations indicate that their hopes for genAI were not always met. While some were excited by its creativity and friendliness, others were disappointed by its lack of cultural references, emotional depth, or engaging responses.

Children’s evaluation of genAI was also closely linked to their fears about over-reliance and its potential academic consequences. When assessing genAI’s clarity, children critiqued how information was presented and expressed the importance of step-by-step explanations that facilitated guided problem-solving rather than simply providing direct answers. For example, Theo articulated this need, stating, ``\textit{Explain how to do it step by step, then I can start working on my homework.}'' Similarly, Jude, when encountering difficulty with a division problem (457 ÷ 9), noted, ``\textit{I [want] a refresher on how to do long division rather than just the exact answer, because that would be cheating.}'' Additionally, children also found that excessively verbose AI-generated responses impeded comprehension. Hazel critiqued the clarity of genAI's output, explaining, ``\textit{It was giving so much information, so it was hard to understand because it gave a six-paragraph response,}'' while Clara echoed this sentiment, stating, ``\textit{it generates long paragraphs that were very intimidating.}'' Ultimately, the process of evaluating genAI encouraged children to reflect on what they value in digital interactions and how technology could be designed to better support their needs.

\section{Discussion \& Future Work}
Children’s perceptions of genAI are shaped by a complex interplay of curiosity, excitement, and concern. While children expressed optimism about genAI’s potential to help them in their everyday lives, they also recognized the uncertainties and risks it presents. This duality highlights a crucial challenge: \textit{how can genAI systems be designed to engage with both their hopes and fears in a meaningful way?} Rather than suppressing fears or amplifying enthusiasm, designing genAI-enabled technologies for children should acknowledge these tensions to equip children with the tools to navigate their evolving relationships with AI \citep{rubegni2022don}. Encouraging thoughtful engagement with genAI, rather than passive adoption, can foster resilience, critical inquiry, and a more nuanced understanding of genAI’s role in children's lives. 

A key finding from our study is that while children envision genAI as a companion, a collaborator, and a task automator, they often view genAI-assisted schoolwork as a high-risk behavior, associating it with punitive consequences rather than as a learning aid. Additionally, while children acknowledged fears about over reliance, their evaluation criteria did not reflect an emphasis on human control or autonomy. Instead, their criteria prioritized genAI’s immediate usability, indicating that while children recognize the risks of dependence, they may not yet possess the critical AI literacy skills necessary to evaluate these risks meaningfully. Thus, as genAI becomes increasingly integrated into educational settings \citep{roose2023chatgpt, grover, chatscratch, alasadi2023generative}, the challenge lies not only in setting clear guidelines for responsible genAI use but also in ensuring that children develop the AI literacy skills needed to critically evaluate, communicate, and collaborate with AI in their daily lives \citep{duri}. In future studies, we aim to explore the agency of children in child-AI collaboration to understand the extent to which children want and need assistance from genAI. Additionally, since our themes are based on observations of 37 children in two fifth-grade classrooms from a single school, our future goal is to explore the hopes and fears across a more diverse group of children to assess the robustness of these themes. Furthermore, while this study focused solely on children’s perspectives, we plan to expand this work to include parents and educators, as their insights could provide valuable perspectives on their hopes, concerns, and experiences with children's use of genAI.

\begin{acks}
This material is based upon work supported under the AI Research Institutes program by the National Science Foundation and the Institute of Education Sciences, U.S. Department of Education, through Award \#DRL-2229873 - AI Institute for Transforming Education for Children with Speech and Language Processing Challenges (or~National AI Institute for Exceptional Education). Any opinions, findings, and conclusions or recommendations expressed in this material are those of the author(s) and do not necessarily reflect the views of the National Science Foundation, the Institute of Education Sciences, or the U.S. Department of Education. This work was also partially funded by the Jacob's Foundation CERES Network.
\end{acks}

\bibliographystyle{ACM-Reference-Format}
\bibliography{references}

\end{document}